\documentstyle[12pt,aaspp4]{article}
\begin{document}

\title{Distance to M31 With the {\em HST}\/ and {\em Hipparcos}\/ 
Red Clump Stars}

\author{K. Z. Stanek \& P. M. Garnavich}
\affil{Harvard-Smithsonian Center for Astrophysics, 60 Garden St., MS20,
Cambridge, MA 02138}
\affil{e-mail: kstanek@cfa.harvard.edu, pgarnavich@cfa.harvard.edu}

\begin{abstract}

Following an approach by Paczy\'nski \& Stanek we compare red clump
stars with parallaxes known to better than 10\% in the {\em
Hipparcos}\/ catalog with the red clump stars observed in three fields
in M31 using the {\em HST}. There are $\sim 600$ and $\sim 6,300$ such
stars in the two data sets, respectively. The local red clump
luminosity function is well represented by a Gaussian with the peak at
$M_{I,m}=-0.23$, and the dispersion $\sigma_{RC}\approx0.2\;$mag.
This allows a single step determination of the distance modulus to M31
$\mu_{0,M31} = 24.471\pm 0.035 \pm 0.045 \;$mag (statistical plus
systematic error) and the corresponding distance $R_{M31}= 784\pm
13\pm 17\;kpc$. The number of red clump stars is large enough that the
formal statistical error in the distance is only $\lesssim 2$\%.

We also correct the treatment of the local interstellar extinction by
Paczy\'nski \& Stanek and we obtain the Galactocentric distance
modulus $\mu_{0,GC}=14.57 \pm0.04 \pm 0.04\;$mag (statistical plus
systematic error), and the corresponding Galactocentric distance
$R_0=8.2 \pm0.15 \pm 0.15\;kpc$.

\end{abstract}

\keywords{Galaxy: center --- galaxies: distances and redshifts ---
galaxies: individual (M31) --- Solar neighborhood --- stars:
horizontal-branch}

\section{INTRODUCTION}

The distance modulus to the M31 galaxy is $\mu_{0,M31} \approx 24.4
\pm0.15\;$mag (for discussion see e.g.~Huterer, Sasselov \& Schechter
1995 and Holland 1998).  In this paper we follow the approach of
Paczy\'nski \& Stanek (1998; hereafter: P\&S) and present an estimate
of the distance to M31 based on the comparison between the red clump
giants observed locally by the {\em Hipparcos}\/ (Perryman et
al.~1997) satellite and observed in M31 with the {\em HST}\/ (Holland,
Fahlman \& Richer 1996; Rich et al.~1996).  These stars are the metal
rich equivalent of the better known horizontal branch stars, and
theoretical models predict that their absolute luminosity only weakly
depends on their age and chemical composition (Seidel, Demarque, \&
Weinberg 1989; Castellani, Chieffi, \& Straniero 1992; Jimenez, Flynn,
\& Kotoneva 1998).  Indeed the absolute magnitude-color diagram of
{\em Hipparcos}\/ (Perryman et al.~1997, their Figure~3) clearly shows
how compact the red clump is.  In this paper we determine the variance
in the $I$-band magnitude to be only $\sim 0.15\;$mag.

As discussed by P\&S, any method of the
distance determination which is based on stars suffers from at least
four problems:

\begin{enumerate}

\item The accuracy depends on the absolute magnitude determination for
the nearby stars;

\item Interstellar extinction has to be determined for the stars in
the target source as well as for those near the Sun;

\item The masses, ages, and chemical composition may be different for
the stars in the source and for their counterparts near the Sun;

\item The statistical error is large if the number of stars is small.

\end{enumerate}

The red clump giants are the only type of stars which do not suffer
from the fourth problem.  In spite of their large number and sound
theoretical understanding these stars have seldom been used as the
distance indicators.  However, recently Stanek (1995) and Stanek et
al.~(1994, 1997) used these stars to map the Galactic bar.  P\&S used
the red clump stars observed by OGLE (Udalski et al.~1993) to obtain
an estimate of the distance to the Galactic center. In this paper we
follow the approach of P\&S and compare the absolute magnitudes of
$\sim 600$ nearby red clump stars with accurate (better than 10\%)
trigonometric parallaxes measured by {\em Hipparcos}\/ with the
apparent magnitudes of $\sim 6,300$ red clump stars observed by the
{\em HST}\/ in the halo of M31 (Holland et al.~1996) and in the M31
globular cluster G1 (Rich et al.~1996).  This comparison gives the
distance to M31 in a single step.

\section{THE DATA AND PRELIMINARY RESULTS}

Inspection of the color-magnitude diagrams based on {\em Hipparcos}\/
and OGLE data revealed a strong dependence of the $V$-band magnitude
of red clump giants on their color, while their $I$-band magnitudes
revealed no significant color dependence (P\&S, their Figures~1,2).
Thus, on purely observational grounds, the $I$-band seems to be the
best in the applications in which the red clump stars are used as
standard candles.  It is possible that bolometric corrections to the
$I$-band are very small for these moderately cool stars, and
theoretical models show only weak dependence of $M_{bol}$ on either
age or chemical composition (Seidel, Demarque, \& Weinberg 1989;
Castellani, Chieffi, \& Straniero 1992; Jimenez, Flynn, \& Kotoneva
1998).

\begin{figure}[t]
\plotfiddle{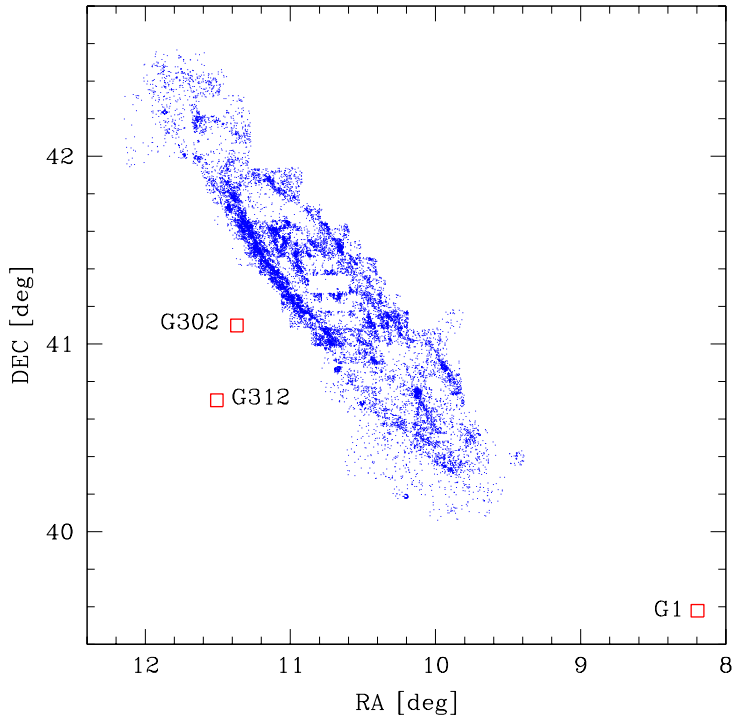}{8cm}{0}{50}{50}{-160}{-90}
\caption{Fields observed by {\em HST}\/ (large rectangles) and used in this
paper to derive the red clump based distance to M31.  Also shown are
blue stars $[(B-V)<0.4]$ from the photometric survey of M31 by Magnier
et al.~(1992) and Haiman et al.~(1994).}
\label{fig:m31}
\end{figure}

\begin{figure}[t]
\plotfiddle{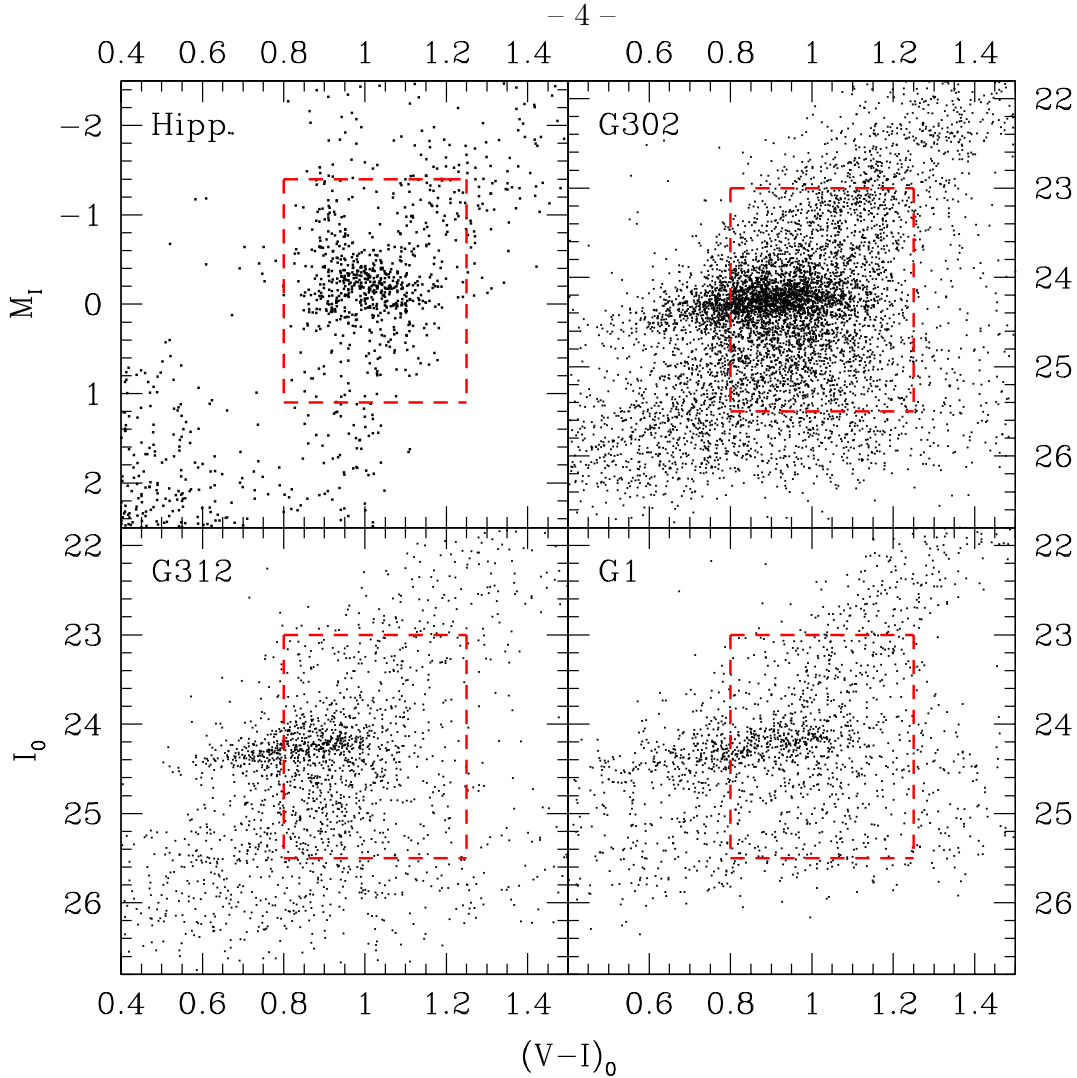}{12.2cm}{0}{70}{70}{-230}{-115}
\caption{The red clump dominated parts of CMDs for the {\em Hipparcos}\/
stars (upper-left panel) and for three fields in M31 observed with the
HST. The dashed rectangles surround the red clump regions used for
the comparison between the local and the M31 red clump stars.}
\label{fig:cmd}
\end{figure}

The {\em Hipparcos}\/ based absolute magnitude-color diagram is shown
in the upper left panel of Figure~\ref{fig:cmd}, for the stars with
parallaxes measured to better than 10\%.  There are 664 such stars
within the dashed rectangle $[0.8<(V-I)<1.25]$ in the upper left panel
of Figure~\ref{fig:cmd}.  P\&S derived the absolute magnitude of the
nearby red clump stars $M_{I,m}=-0.185 \pm0.016$. We will discuss this
number later in the paper.

Deep CMDs were obtained for several lines-of-sight towards M31, mostly
to study the M31 globular clusters (Ajhar et al.~1996; Fusi Pecci et
al.~1996; Holland et al.~1996, 1997; Rich et al.~1996). We selected
the $V,\;V-I$ CMDs obtained by Rich et al.~(1996) for the M31 globular
cluster G1 and by Holland et al.~(1996) for two fields (adjacent to
the M31 globular clusters G302 and G312) in the M31 halo
(Figure~\ref{fig:m31}). These fields are located $\sim 30\arcmin,
50\arcmin$ and $150\arcmin$ from the center of M31, respectively. The
G302 and G312 CMDs were corrected for the reddening and extinction
using a value of $E(B-V)=0.08$ (Burstein \& Heiles 1982).  For G1, we
used a value of $E(B-V)=0.058$ taken from Schlegel, Finkbeiner \&
Davis (1998).  The red-clump dominated parts of the G302, G312 and G1
CMDs are shown in the upper-right, lower-left and lower-right panels
of Figure~\ref{fig:cmd}, respectively.  As discussed by Holland et
al.~(1996), CMDs for G302 and G312 indicate multiple stellar
populations, consistent with a mix of 50\% to 75\% metal-rich stars
and 25\% to 50\% metal-poor stars.  This is reflected by the presence
of horizontal branch stars in M31 bluer $[(V-I)_0<0.8]$ than the red
clump stars in the Solar neighborhood. Rich et al.~(1996) concluded
that the properties of the G1 CMD are most consistent with those of an
old globular cluster with the metallicity of 47 Tuc.  Given the fact
that the average luminosity of the red clump stars appears to be
independent of their color, and hence metallicity, in Baade's Window
(P\&S, their Figure~1) and in the Solar neighborhood
(Figure~\ref{fig:cmd}, upper-left panel), it seems safe to determine
the distance to M31 by comparing these two populations, especially
using the overlapping region in the color.

\begin{figure}[t]
\plotfiddle{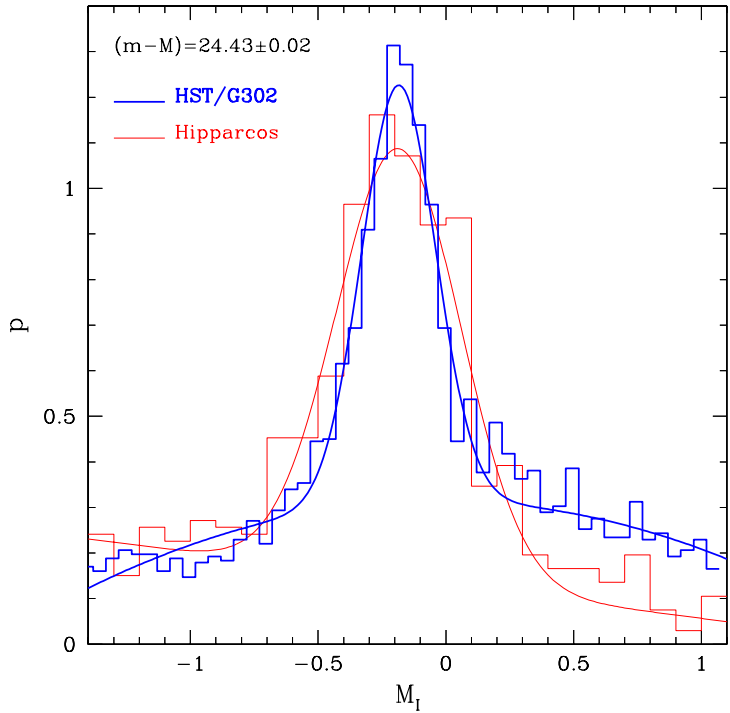}{8cm}{0}{50}{50}{-160}{-90}
\caption{The number of red clump stars in the solar neighborhood,
based on {\em Hipparcos}\/ data, is shown as a function of absolute
magnitude $M_I$ with the thin solid line (fit and histogram). The
number of red clump stars in the G302 field (Holland et al.~1996), is
shown as a function of absolute magnitude $M_I$ with thick solid line,
adopting the preliminary distance modulus: $I_0 - M_I = 24.43$.  All
distributions are normalized.}
\label{fig:dist}
\end{figure}

Following P\&S, we selected the red clump stars in the color range
$0.8<(V-I)_0<1.25$ in the G302 field (4357 stars), in the G312 field
(980 stars) and in the G1 globular cluster (937 stars).  The color
range was selected to correspond to the color range of the red clump
stars observed locally by the {\em Hipparcos} (Figure~\ref{fig:cmd},
upper-left panel). We fitted all three distributions with a function
\begin{equation}
n(I_0) = a + b (I_0-I_{0,m})  + c (I_0-I_{0,m})^2 +
\frac{N_{RC}}{\sigma_{RC}\sqrt{2\pi}}
 \exp\left[-\frac{(I_0-I_{0,m})^2}{2\sigma_{RC}^2} \;\right]
\end{equation}
(please note the typo in P\&S Eq.1, where the factor $2$ in the
denominator of the exponent was, incorrectly, squared). The first
three terms describe a fit to the ``background'' distribution of the
red giant stars, and the Gaussian term represents a fit to the red
clump itself.  $I_{0,m}$ correspond to the peak magnitude of the red
clump population. We obtained the values of $I_{0,m}=24.245 \pm0.006$
for the G302 field, $I_{0,m}=24.222 \pm0.014$ for the G312 field and
$I_{0,m}=24.209\pm0.013$ for the G1 cluster. This, in combination with
the $M_{I,m}=-0.185 \pm0.016$ for the local red clump stars, gives
preliminary distance moduli of $(m-M)_{G302}=24.43 \pm0.02$,
$(m-M)_{G312}=24.41 \pm0.02$ and $(m-M)_{G1}=24.39 \pm0.02$.

\begin{figure}[t]
\plotfiddle{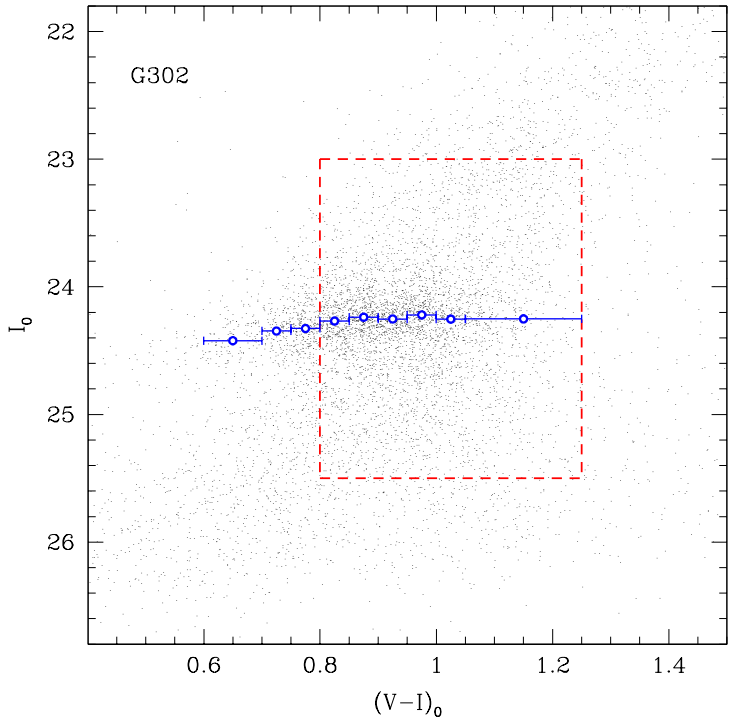}{8cm}{0}{50}{50}{-160}{-90}
\caption{Dependence of the peak brightness of the red clump $I_{0,m}$
on the $(V-I)_0$ color for the most star-rich M31 field G302. The red
clump exhibits a relatively sharp downturn for $(V-I)_0<0.8\;$mag, but
there is no color dependence for $(V-I)_0>0.8\;$mag.  In the color
range $0.8<(V-I)_0<1.25$, used for the comparison with the {\em
Hipparcos}\/ red clump, $I_{0,m}$ varies only from 24.22 to 24.268 and
in a random fashion.}
\label{fig:morph}
\end{figure}

The distribution of the local and the M31/G302 red clump stars as a
function of their absolute $I$-band magnitude is shown in
Figure~\ref{fig:dist} with two histograms as well as with two
analytical fits of the type described by the Eq.1.  All distributions
are normalized.  The Gaussian fitted to the G302 field red clump
distribution has a small $\sigma_{RC}= 0.15\;mag$ and the Gaussian
fitted to the {\em Hipparcos}\/ distribution has $\sigma_{RC}=
0.24\;mag$. This difference could be due to the {\em Hipparcos}\/
stars having $\lesssim$10\% parallax errors, broadening the derived
distribution. By the same argument, the narrowness of the red clumps
in M31 indicates that the {\em HST} $I$-band photometry is still very
accurate at $I\sim24.5\;$mag.

As we discussed above, P\&S found no significant color dependence in
the $I$-band magnitudes of the red clump stars in both the {\em
Hipparcos}\/ and the OGLE data.  However, a visual examination of the
M31 data in Figure~\ref{fig:cmd} suggests that the blue edge of the
red clump is $\sim 0.2\;$mag fainter than the red edge. To investigate
this effect, we divided the red clump region of the most star-rich
G302 field into nine color bins and for each bin we performed the fit
described above. There were between 250 and 1,050 stars in each
bin. The result appears in Figure~\ref{fig:morph}, where we show
$I_{0,m}$ for each bin. The red clump exhibits a relatively sharp
downturn for $(V-I)_0<0.8\;$mag, but there is basically no color
dependence for $(V-I)_0>0.8\;$mag, consistent with the result of P\&S.
In the color range $0.8<(V-I)_0<1.25$, used for the comparison with
the {\em Hipparcos}\/ red clump, $I_{0,m}$ varies only from 24.22 to
24.268 and in a random fashion.

\section{CORRECTING SOME ERRORS}

The very small formal statistical error for our distance to M31
follows from the very large number of red clump stars measured by {\em
Hipparcos}\/ and observed in M31.  The number of these stars is
several orders of magnitude larger than the number of either RR Lyrae
or Mira variables, and the observed dispersion in their magnitudes is
as small as $\sigma_{RC}\approx0.15\;mag$ in the {\em HST}\/ data for
the M31 fields.  However, in addition to small statistical errors
there are possibly larger systematic errors discussed by P\&S, some of
which we try to estimate and correct for.

P\&S checked if the 10\% error limit adopted was not too generous, and
they repeated the analysis with a more stringent 5\% upper limit to
the parallax errors.  This reduced the number of stars within the
rectangle shown in Figure~\ref{fig:cmd} from 664 to 240, and gave
$M_{I,m} = -0.173 \pm 0.027$, within statistical errors of $M_{I,m} =
-0.185 \pm 0.016$ obtained for the 10\% sample.

By selecting stars with accurate parallaxes from the {\em Hipparcos}\/
catalog we introduced a distance bias which depends on the absolute
magnitude: the stars which are intrinsically brighter can be measured
accurately out to a larger distance.  Therefore, there are relatively
more bright stars in our {\em Hipparcos}\/ sample than in a
volume-limited sample.  In order to estimate this effect we selected a
sub-sample of 228 {\em Hipparcos}\/ stars with the distance
$d<70\;pc$, and we determined the parameters of the best fit to the
luminosity function described with Eq.1.  This sample is not strictly
speaking volume limited, but closer to that than the previous sample.
We found that $M_{I,m}(d<70\;pc) = -0.227 \pm 0.023$, and
$\sigma_{RC}=0.209$ mag for these nearby stars, with the average
distance $\langle d_{<70}\rangle =50\;pc$. As discussed by H${\rm
\o}$g \& Flynn (1998, their Figure~2), one expects very little
reddening, $E(B-V)<0.02$, for such nearby stars, so we assume that our
$d<70\;pc$ sample suffers no reddening. This is contrary to what was
assumed by P\&S, but agrees with detailed models of the optical
reddening in the Solar neighborhood (M\'endez \& van Altena 1998): we
live in a bubble of low interstellar extinction (Bhat, Gupta \& Rao
1998).  We therefore assume that a Gaussian with $M_{I,m} =
-0.23\pm0.03$ and dispersion $\sigma_{RC}\approx 0.2\;$mag well
represents the red clump luminosity function. We believe that a
$1\sigma$ error of $0.03\;$mag better represents the uncertainty of
the red clump properties than the value of $0.09\;$mag derived by
P\&S.  We correct the value of the Galactocentric distance modulus
obtained by P\&S to $(M-m)_{GC}=14.57 \pm0.04 \pm 0.04\;$mag
(statistical plus systematic error), and the corresponding distance
$R_0=8.2 \pm0.15 \pm 0.15 \;kpc$. The systematic error combines the
uncertainty of $0.025\;$mag in the $I$-band zero point of the Stanek
(1996) extinction map (Gould, Popowski \& Terndrup 1998; Alcock et
al.~1998) as well as possible $0.03\;$mag error in the OGLE $I$-band
photometric zero point (Kiraga, Paczy\'nski \& Stanek 1997).

There are also a number of systematic errors connected with the M31
halo red clump stars.  The G1 globular cluster might be at a distance
that is different from the bulk of M31. Similarly, the centroid of the
Gaussian distribution of red clump giants in the M31 halo corresponds
to the distance which might differ from the distance to the disk of
M31.  However, as discussed by Holland et al.~(1996), the depth of the
M31 halo is very small ($\sim0.02\;$mag), so the possible shift in the
M31 distance is even smaller.  This is confirmed by the very similar
values of distance moduli we find for the two lines-of-sight in the
halo of M31.  A larger contribution to the error comes from the zero
point in the assumed interstellar extinction, which Holland et
al.~(1996) took to be $E(B-V)=0.08$ (Burstein \& Heiles 1982), but
Schlegel et al.~(1998) advocate a lower value of $E(B-V)=0.062$ as an
average foreground reddening for M31. However, using the Schlegel et
al.~(1998) reddening map directly at the positions of G302 and G312
gives $E(B-V)=0.084$ and $E(B-V)=0.082$, respectively.  For G1 we
obtained a value of $E(B-V)=0.058$ using the map of Schlegel et
al.~(1998), in excellent agreement with value $E(B-V)=0.06$ adopted
by Rich et al.~(1996). Based on discussion by Schlegel et al.~(1998)
and Stanek (1998), we estimate the $1\sigma$ systematic error in the
$E(B-V)$ reddening to be $\sim0.015\;$mag, which using the relation
$A_I/E(B-V)=1.95$ translates to systematic $I$-band extinction error
of $0.03\;$mag.

Another systematic error is the uncertainty in the WFPC2 photometric
zero point thought to be as high as $0.05\;$mag based on calibrations
by the Distance Scale Key Project (Hill et al.~1998).  Recently
Whitmore \& Heyer (1997) showed that the charge transfer efficiency
(CTE) for the WFPC2 CCDs depended on the total number of electrons in
the star image and the sky background as well as the more commonly
known problem with position on the detector. Corrections for the full
CTE effect would reduce the zero point uncertainty to $0.03\;$mag, but
only the standard Holtzman et al.~(1995) ramp was applied to the data
presented here.  Fortunately the red clump stars are relatively
bright, and the difference between the simple ramp CTE correction and
the Whitmore correction amounts to only 2\%. We therefore assume that
the systematic error due to the WFPC2 photometric zero point is
$0.035\;$mag.

We can now derive the final distance modulus to M31 based on the
lines-of-sight discussed in this paper.  The straight average of red
clump peak apparent magnitudes $I_{0,m}$ for the three fields is
$\langle I_{0,m} \rangle = 24.225 \pm 0.018$ and the weighted mean is
$\bar{I}_{0,m} = 24.236 \pm 0.021$. However, as pointed out to us by
the referee, if we assume that the M31 GC system is dynamically
similar to the Galactic GC system, then it is highly probable that G1
does not lie at the same distance as M31 but could be expected to have
a distance modulus of $\sim 0.035\;$mag greater or smaller than that
of M31, due to G1's location in its orbit about M31.  This appears to
be consistent with G1 having $I_{0,m}$ $0.025\;$mag less than the mean
$I_{0,m}$ of the G302 and G312 fields (however, the difference is
small enough, that it could be caused by, for example, slight
uncertainty in the {\em HST}\/ photometric zero point discussed
above). Using only the G302 and G312 field stars, the straight average
of red clump peak apparent magnitudes $I_{0,m}$ is $\langle I_{0,m}
\rangle = 24.233 \pm 0.016$ and the weighted mean is $\bar{I}_{0,m} =
24.241 \pm 0.016$. Using this weighted mean $\bar{I}_{0,m}$ combined
with the distribution of local red clump stars, we obtain the distance
modulus to M31 $\mu_{0,M31} = 24.471 \pm 0.035\;$mag or $R_{M31}=784
\pm 13\;kpc$ (statistical error only). After adding to that the
systematic error of $0.03\;$mag due to the uncertainty in $E(B-V)$
determination and $0.035\;$mag due to the zero-point uncertainty in
the {\em HST}\/ photometry, we arrive at the final value of
$\mu_{0,M31}= 24.471 \pm 0.035 \pm 0.045\;$mag (statistical plus
systematic error). It is worth noticing that Freedman \& Madore (1990)
obtained very close average value of $\mu_{0,M31}= 24.44 \pm
0.13\;$mag for a sample of Cepheids observed in Baade's Fields I, III
and IV (Baade \& Swope 1963, 1965).

There is yet another type of a systematic error possible: the age, the
chemical composition, and the masses of red clump giants may be
systematically different in the M31 halo and near the Sun. As
discussed in the previous section, the presence of horizontal branch
stars with $(V-I)_0<0.8$ in the halo of M31, which are basically
absent in the {\em Hipparcos}\/ sample, implies that the two
populations are to some extent different (see Holland et al.~1996 and
Rich et al.~1996 for more detailed discussions).  Recent stellar
evolutionary models (Jimenez et al.~1998) indicate that in the age
range 2--12 billion years the effective temperature is dominated by
the metallicity. As we discussed earlier, by comparing the red clump
distributions selected using the same color range we hope to minimize
the impact of the population differences on the derived distance
modulus of M31.  As seen in Figure~\ref{fig:m31}, our three
lines-of-sight probe a large range of M31 galactocentric distances and
locations (two fields along the SE minor axis and one along the SW
major axis) and hence metallicities and possibly ages and star
formation histories. The fact that the derived distance moduli for the
three fields we used in this paper vary by only $\sim 0.035\;$mag
indicates that indeed our approach is valid.

To summarize, among the various stellar distance indicators the red
clump giants might be the best for determining the distance
to M31 because there are so many of them.  In particular {\em
Hipparcos}\/ provided accurate distance determinations for almost
2,000 such stars, but unfortunately $I$-band photometry is available
for only $\sim 30\%$ of them, so it would be important to obtain
$I$-band photometry for all {\em Hipparcos}\/ red clump giants. One
would also like to use many more lines-of-sight towards M31 to
understand better the effects of varying stellar populations and
reduce the uncertainty in the {\em HST}/{\em Hipparcos}\/ comparison.

\paragraph{NOTE:} After we completed writing this paper we became
aware of the paper by Holland (1998), in which he determines the
distance to M31 using red-giant branches of globular clusters in
M31. His distance modulus of $\mu_0=24.47\pm0.07\;$mag is identical
to the value of $\mu_{0,M31}= 24.471 \pm 0.035 \pm 0.045\;$mag we
obtained in this paper.

\acknowledgments{KZS was supported by the Harvard-Smithsonian Center
for Astrophysics Fellowship. We would like to thank Stephen Holland,
Mike Rich and Don Neill for kindly making their {\em HST}
color-magnitude data available to us. It is a great pleasure to
acknowledge helpful comments by Andy Gould and Dimitar Sasselov.  We
also thank the referee for fast and extremely careful reading of the
manuscript and his very useful and detailed comments.}


\begin{references}

\reference{} Ajhar, E. A., et al., 1996, AJ, 111, 1110

\reference{} Alcock, C., et al., 1998, ApJ, 494, 396

\reference{} Baade, W., \& Swope, H. H., 1963, AJ, 68, 435

\reference{} Baade, W., \& Swope, H. H., 1965, AJ, 70, 212 

\reference{} Bhat, N. D. R., Gupta, Y., \& Rao, A. P., 1998, ApJ, in press
	(astro-ph/9802203)

\reference{} Castellani, V., Chieffi, A., \& Straniero, O., 1992, ApJS, 78, 517

\reference{} Freedman, W. L., \& Madore, B. F., 1990, ApJ, 365, 186 

\reference{} Fusi Pecci, F., et al., 1996, AJ, 112, 1461

\reference{} Gould, A., Popowski, P., \& Terndrup, D. M., 1998, ApJ, 492, 778

\reference{} Haiman, Z., et al., 1994, A\&A, 286, 725

\reference{} Hill, R. J., et al. ({\em HST Key Project Team}), 1998,
	ApJ, in press

\reference{} Holland, S., Fahlman, G. G., \& Richer, H. B., 1996, AJ, 112, 1035

\reference{} Holland, S., Fahlman, G. G., \& Richer, H. B., 1997, AJ, 114, 1488

\reference{} Holland, S., 1998, AJ, to appear in May 1998 issue 
	(astro-ph/9802088)

\reference{} Holtzman, J. A., et al., 1995, PASP, 107, 1065

\reference{} H${\rm \o}$g, E., et al., 1997, A\&A, 323, L57

\reference{} H${\rm \o}$g, E., \& Flynn, C., 1998, MNRAS, in press
	(astro-ph/9708061)

\reference{} Huterer, D., Sasselov, D. D., \& Schechter, P. L., 
	1995, AJ, 110, 2705

\reference{} Jimenez, R., Flynn, C., \& Kotoneva, E. 1998, MNRAS,
	in press (astro-ph/9709056)

\reference{} Kiraga, M., Paczy\'nski, B., \& Stanek, K. Z., 1997, ApJ, 485, 611

\reference{} Magnier, E. A., et al., 1992, A\&AS, 96, 37

\reference{} M\'endez, R. A., \& van Altena, W. F., 1998, A\&A, in press
	(astro-ph/9710030)

\reference{} Paczy\'nski, B., \& Stanek, K. Z., 1998, ApJ, 494, L219 [P\&S]

\reference{} Perryman, M. A. C., et al., 1997, A\&A, 323, L49

\reference{} Rich, R. M., Mighell, K. J., Freedman, W. L., \& Neill, J. D., 
  	1996, AJ, 111, 768

\reference{} Seidel, E., Demarque, P., \& Weinberg, D. 1987, ApJS, 63, 917

\reference{} Schlegel, D. J.,  Finkbeiner, D. P., \&  Davis, M., 1998, ApJ,
	in press (astro-ph/9710327)

\reference{} Stanek, K. Z., et al., 1994, ApJ, 429, L73

\reference{} Stanek, K. Z., 1995, ApJ, 441, L29

\reference{} Stanek, K. Z., 1996, ApJ, 460, L37

\reference{} Stanek, K. Z., et al., 1997, ApJ, 477, 163

\reference{} Stanek, K. Z., 1998, ApJ, submitted (astro-ph/9802093)

\reference{} Udalski, A., Szyma\'nski, M., Ka\l u\.zny, J., Kubiak, M.,
        \& Mateo, M., 1993, AcA, 43, 69

\reference{} Whitmore, B., \& Heyer, I., 1997, {\em HST}\/ Instrument Science 
	Report, WFPC2 97-08


\end{references}
\end{document}